A Detailed Characterization of the Expert Problem-Solving Process in Science and Engineering; Guidance for Teaching and Assessment

**Manuscript type:** Article

**Shortened title:** Characterization of Expert Problem-Solving

**Authors**

Argenta M. Price[1][*], Candice J. Kim[2,3], Eric W. Burkholder[1], Amy V. Fritz[4], and Carl E. Wieman[1,2]

**Affiliations**

[1] Stanford University, Department of Physics, Stanford, CA 94305.

[2] Stanford University, Graduate School of Education, Stanford, CA 94305.

[3] Stanford University, School of Medicine, Stanford, CA 94305.

[4] Stanford University, Department of Electrical Engineering, Stanford, CA 94305.

[*] Corresponding Author

**Keywords:** expertise, problem-solving, undergraduate science education, graduate science education, theoretical framework

**Abstract**

A primary goal of science and engineering (S & E) education is to produce good problem solvers, but how to best teach and measure the quality of problem-solving remains unclear. The process is complex, multifaceted, and not fully characterized. Here we present a theoretical framework of the S & E problem-solving process as a set of specific interlinked decisions. This theory is empirically grounded and describes the entire process. To develop this theory, we interviewed 52 successful scientists and engineers ("experts") spanning different disciplines, including biology and medicine. They described how they solved a typical but important problem in their work, and we analyzed the interviews in terms of decisions made. Surprisingly, we found that across all experts and fields, the solution process was framed around making a set of just twenty-nine specific decisions. We also found that the process of making those discipline-general decisions (selecting between alternative actions) relied heavily on domain-specific predictive models that embodied the relevant disciplinary knowledge. This set of decisions provides a guide for the detailed measurement and teaching of S & E problem-solving. This decision framework also provides a more specific, complete, and empirically based theory describing the "practices" of science.

**Introduction**

Many faculty members with new graduate students and many managers with employees who are recent college graduates have had similar experiences. Their advisees/employees have just completed a program of rigorous course work, often with distinction, but they seem unable to solve the real-world problems they encounter. The supervisor struggles to figure out exactly what the problem is and how they can guide the person in overcoming it. This paper is providing a way to answer those questions in the context of science and engineering.

The importance of problem-solving as an educational outcome has long been recognized, but too often post-secondary science and engineering (S & E) graduates have serious difficulties when confronted with real world problems (QS, 2018). This reflects two long standing educational problems with regard to problem-solving: how to properly measure it, and how to effectively teach it. We theorize that the root of these difficulties is that good "problem-solving" is a complex multifaceted process, and the details of that process have not been sufficiently characterized. Better characterization of the problem-solving process is necessary to allow problem-solving, and more particularly, the complex set of skills and knowledge it entails, to be measured and taught more effectively. We sought to create an empirically grounded theoretical framework that would characterize the detailed structure of the full problem-solving process used by skilled practitioners when solving problems as part of their work. We also wanted a framework that would allow use and comparison across S & E disciplines. To create such a framework, we examined the operational decisions (those that result in subsequent actions) that these practitioners make when solving problems in their discipline.

Various aspects of problem-solving have been studied across multiple domains and using a variety of methods (e. g. Newell & Simon, 1972; Dunbar, 2000; NRC, 2012a; Lintern et al., 2018). These range from expert self-reflections (e. g. Polya, 1945), to studies on knowledge-lean-tasks to discover general problem-solving heuristics (e. g. Egan & Greeno, 1974), to comparing expert and novice performance on simplified problems across a variety of disciplines (e. g. Ericsson et al., 2006, 2018; Chase & Simon, 1973; Chi et al., 1981; Larkin & Reif, 1979). These studies revealed important novice-expert differences – notably, that experts are better at identifying important features, and have knowledge structures that allow them to reduce demands on working memory. Studies that specifically gave the experts unfamiliar problems in their disciplines also found that, relative to novices, they had more deliberate and reflective strategies, including more extensive planning and managing of their own behavior, and they could use their knowledge base to better define the problem (Schoenfeld, 1985; Wineburg, 1988; Singh, 2002). While this approach focused on discrete cognitive steps of the individual, an alternative framing of problem solving has been in terms of "ecological psychology" of "situativity", looking at how the problem solver views and interacts with their environment in terms of affordances and constraints (Greeno, 1994). "Naturalistic decision making" is a related framework that specifically examines how experts make decisions in complex, real-world, settings, with an emphasis on the importance of assessing the situation surrounding the problem at hand (Mosier et al., 2018; Klein, 2008).

While this work on expertise has provided important insights into the problem-solving process, its focus has been limited. Most has focused on looking for cognitive differences between experts and novices using limited and targeted tasks, such as remembering the pieces on a chess board (Chase & Simon, 1973) or identifying the important concepts represented in an introductory physics textbook problem (Chi et al., 1981). It did not attempt to explore the full process of solving, particularly for solving the type of complex problem that a scientist or engineer encounters as a member of the workforce ("authentic problems").

There have also been many theoretical proposals as to expert problem-solving practices, but with little empirical evidence as to their completeness or accuracy (e.g. Polya, 1945; Heller & Reif, 1984; OECD 2019). The work of Dunbar is a notable exception to the lack of empirical work, as his group did examine how biologists solved problems in their work by analyzing lab meetings held by eight molecular biology research groups (2000). His ground-breaking work focused on creativity and discovery in the research process, and he identified the importance of analogical reasoning and distributed reasoning by scientists in their answering research questions and gaining new insights. Kozma et al. (2000) studied professional chemists solving problems, but their work focused only on the use of specialized representations.

The "cognitive systems engineering" approach (Lintern et al., 2018) takes a more empirically based approach looking at experts solving problems in their work, and as such tends to span aspects of both the purely cognitive and the ecological psychological theories. It uses both observations of experts in authentic work settings and retrospective interviews about how experts carried out particular work tasks. This theoretical framing and the experimental methods are similar to what we use, particularly in the "naturalistic decision making" area of research (Mosier et al., 2018). That work looks at how critical decisions are made in solving specific problems in their real-world setting. The decision process is studied primarily through retrospective interviews about challenging cases faced by experts. As described below, our methods are adapted from that work (Crandall et al., 2006), though there are some notable differences in focus and field. A particular difference is that we focused on identifying *what are* decisions-to-be-made, which are more straight-forward to identify from retrospective interviews than *how* those decisions are made. We all have the same ultimate goal, however, to improve the training/teaching of the respective expertise.

Problem solving is central to the process of science, engineering, and medicine, so research and educational standards about scientific thinking and the process and practices of science are also relevant to this discussion. Work by Osborne and colleagues describes six styles of scientific reasoning, which can be used to explain how scientists and students approach different problems (2016). There are also numerous educational standards and frameworks that, based on theory, lay out the skills or practices that science and engineering students are expected to master (e.g AAAS 2011, NGSS Lead States, 2013,

ABET 2020, OECD 2019). More specifically related to the training of problem-solving, Priemer et al. (2020) synthesizes literature on problem solving and scientific reasoning to create a "STEM and computer science framework for problem solving" that lays out steps that could be involved in a students' problem solving efforts across STEM fields. These frameworks provide a rich groundwork, but they have several limitations: 1) They are based on theoretical ideas of the practice of science, not empirical evidence, so while each framework contains overlapping elements of the problem-solving process, it is unclear whether they capture the complete process. 2) They are focused on school-science, rather than the actual problem-solving that practitioners carry out and that students will need to carry out in future STEM careers. 3) They are typically under-specified, so that the steps or practices apply generally, but it is difficult to translate them into measurable learning goals for students to practice. Working to address that, Clemmons et al. (2020) recently sought to operationalize the core competencies from the Vision and Change report (AAAS, 2011), establishing a set of skills that biology students should be able to master.

Our work seeks to augment this prior work by building a theoretical framework that is empirically based, grounded in how scientists and engineers solve problems in practice instead of in school. We base our theory on the decisions that need to be made during problem-solving, which makes each item clearly defined for practice and assessment. In our analysis of expert problem-solving, we empirically identified the entire problem-solving process. We found this includes deciding when and how to use the steps and skills defined in the work described above but also includes additional elements. There are also questions in the literature about how generalizable across fields a particular set of practices may be. Here we present the first empirical examination of the entire problem-solving process, and we compare that process across many different S & E disciplines.

A variety of instructional methods have been used to try and teach science and engineering problem-solving, but there has been little evidence of their efficacy at improving problem-solving (see [NRC, 2012a] for a review). Research explicitly on teaching problem-solving has primarily focused on textbook-type exercises and utilized step-by-step strategies or heuristics. These studies have shown limited success, often getting students to follow specific procedural steps but with little gain in actually solving problems, and showing some potential drawbacks (Heller & Reif, 1984; Huffman, 1997; Heller et al., 1992; Heckler, 2010; Kuo et al., 2017). As discussed below, the framework presented here offers guidance for different and potentially more effective approaches to teaching problem-solving.

These challenges can be illustrated by considering three different problems, taken from courses in mechanical engineering, physics, and biology, respectively (Fig. 1). All of these are challenging, requiring considerable knowledge and effort by the student to solve correctly.

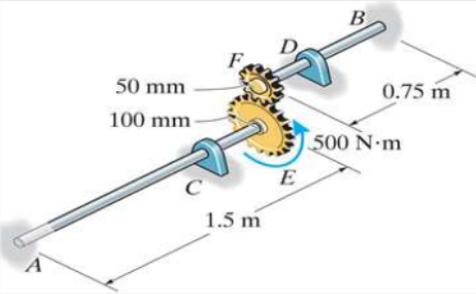

**Mechanical Engineering** *"The two shafts shown in the figure below are made of A-36 steel. Each has a diameter of 25 mm and they are connected using the gears fixed to their ends. Their other ends are attached to fixed supports at A and B. They are also supported by journal bearings at C and D, which allow free rotation of the shafts along their exes. If a torque of 500 N.m is applied to the gear at E as shown in the figure, determine the reactions at A & B. For A-36 Steel: E = 200 GPa and G = 75 GPa."*
From: https://engineering.wayne.edu/me/exams/ thermodynamics_-_sample_pqe_problems_.pdf

**Physics** *"Solve for the eigenvalues and eigenfunctions of the hydrogen atom using the Dirac equation."* A standard physics problem in nearly every advanced quantum mechanics course.

**Biology** *"Comparisons of the patterns of mRNA levels across different human cell types show that the level of expression of almost every active gene is different. The patterns of mRNA abundance are so characteristic of cell type that they can be used to determine the tissue of origin of cancer cells, even though the cancer cells may have metastasized to different parts of the body. By definition, however, cancer cells are different from their noncancerous precursor cells. How do you suppose then that patterns of mRNA expression might be used to determine the tissue source of a human cancer?"* From "Molecular Biology of the Cell", 6th edition, chapter 7 end of chapter problems. (Alberts, 2014)

Figure 1: Example problems from courses or textbooks in mechanical engineering, physics and biology.

Problems such as these are routinely used to both assess students' problem-solving skills, and students are expected to learn such skills by practicing doing such problems. However, it is obvious to any expert in the respective fields, that, while these problems might be complicated and difficult to answer, they are vastly different from solving authentic problems in that field. They all have well-defined answers that can be reached by straightforward solution paths. More specifically, they do not involve needing to use judgement to make any decisions based on limited information, e.g., insufficient to specify correct decision with certainty. The relevant concepts and information and assumptions are all stated or obvious. The failure of problems like these to capture the complexity of authentic problem solving underlies the failure of efforts to measure and teach problem solving. Recognizing this failure motivated our efforts to more completely characterize the problem-solving process of practicing scientists, engineers, and doctors.

We are building on the previous work studying expert-novice differences and problem solving but taking a different direction. We sought to create an empirically grounded framework that would characterize the detailed structure of the full problem-solving process, by focusing on the operational decisions that skilled practitioners make when successfully solving "authentic problems" in their scientific, engineering, or medical work. We chose to identify the decisions that S & E practitioners made, because, unlike potentially nebulous skills or general problem-solving steps that might change with the discipline, decisions are sufficiently specified that they can be individually practiced by students and measured by instructors or departments. The "authentic problems" that we analyzed are typical problems practitioners encounter in "doing" the science or engineering that is their job. In the language of traditional problem solving and expertise research, such authentic problems are "ill-structured" (Simon, 1973) and require "adaptive expertise" (Hatano & Inagaki,1986) to solve. However, our "authentic" problems are considerably more complex and unstructured than what is normally considered in those literatures, because not only do they not have a clear solution path, in many cases it is not clear a priori that they have any solution at all. Determining that, and if the problem needs to be redefined to be soluble, is part of the successful expert solution process. Another way in which our set of decisions goes beyond the characterization of what is involved in adaptive expertise is the prominent role of making judgements with limited information.

A common reaction of scientists and engineers to seeing the list of decisions we obtain as our primary result is, "Oh yes, these are things I always do in solving problems. There is nothing new here." It is comforting that these decisions all look familiar; that supports their validity. However, what is new is not that experts are making such decisions, but rather that there is a relatively small but complete set of decisions that has now been explicitly identified and that apply so generally.

We have used a much larger and broader sample of experts in this work than used in prior expert-novice studies, and we used a more stringent selection criterion. Previous empirical work has typically involved just a few experts, almost always in a single domain, and in some cases included graduate students as "experts". Our sample was 31 experienced practitioners from ten different disciplines of science, engineering, and medicine, with demonstrated competence and accomplishments well beyond that of most graduate students. Also, approximately 25 additional experts from across science, engineering, and medicine served as consultants during the planning and execution of this work.

Our research question was: *What are the decisions experts make in solving authentic problems, and to what extent is this set of decisions-to-be-made consistent both within and across disciplines?*

Our approach was designed to identify the level of consistency and unique differences across disciplines. Our hypothesis was that there would be a manageable number (20-50) of decisions-to-be-made, with a large amount of overlap of decisions made between experts within each discipline, and a substantial but smaller overlap across disciplines. We believed that if we had found that every expert and/or discipline used a large and completely unique set of decisions, it would have been an interesting research result but of little further use. If our hypothesis turned out to be correct, we expected that the set of decisions obtained would have useful applications in guiding teaching and assessment, as they would show how experts in the respective disciplines applied their content knowledge to solve problems and hence provide a model for what to teach. We were not expecting to find the nearly complete degree of overlap in the decisions made across all the experts.

**Methods**

We first conducted 22 relatively unstructured interviews with a range of S & E experts, where we asked about problem-solving expertise in their field. From these interviews, we developed an initial list of decisions-to-be-made in S & E problem-solving. To refine and validate the list, we then carried out a set of 31 semi-structured interviews in which S & E experts chose a specific problem from their work and described the solution process in detail. These interviews were coded for the decisions represented, either explicitly stated or implied by a choice of action. This provided a framework of decisions that characterize the problem-solving process across S & E disciplines. The details of the method are discussed below. The research was approved by the Stanford IRB #48785 and informed consent was obtained from all the participants.

This work involved interviewing many experts across different fields. We defined experts as practicing scientists, engineers, or physicians with considerable experience working as faculty at highly rated universities or having several years of experience working in moderately high-level technical positions at successful companies. We also included a few long-time post-docs and research staff in biosciences to capture more details of experimental decisions from which faculty members in those fields often were more removed. This definition of expert allows us to identify the practices of skilled professionals; we are not studying what makes only the most exceptional experts unique.

Experts were volunteers recruited through direct contact via the research team's personal and professional networks, and referrals from experts in our networks. Within this limitation, we attempted to get a large range of experts by field and experience. This included people from ten different fields (including molecular biology/biochemistry, ecology, and medicine), eleven US universities, and nine different companies or government labs, and was 33% female. The medical experts were volunteers from a select group of medical school faculty chosen to serve as clinical reasoning mentors for medical students at a prestigious university. We only contacted people that met our criteria for being an "expert", and everyone that volunteered was included in the study. Most of the people who were contacted volunteered, and the only reason given for not volunteering was insufficient time. Other than their disciplinary expertise, there was little to distinguish these experts beyond the fact they were acquaintances with members of the team or acquaintances of acquaintances of team or project advisory board members. The precise number from each field was determined largely by availability of suitable experts.

We defined an "authentic problem" to be one that these experts solve in their actual jobs. Generally, this meant research projects for the science and engineering faculty, design problems for the industry engineers, and patient diagnoses for the medical doctors. Such problems are characterized by complexity, with many factors involved and no obvious solution process, and involving substantial time, effort, and resources. Such problems involve far more complexity and many more decisions, particularly decisions with limited information, than the typical problems used in previous problem-solving research or used with students in instructional settings.

**Creating initial list of problem-solving decisions**

We first interviewed 22 experts (Table 1), most of whom were faculty at a prestigious university, in which we asked them to discuss expertise and problem-solving in their field as it related to their own experience. This usually resulted in their discussing examples of one or more problems they had solved.
Based on the first seven interviews, plus reflections from the research team and review of the literature
on expert problem-solving and teaching of scientific practices (Ericsson et al., 2006; Wieman, 2015; NRC,
2012b) , we created a generic list of decisions that were made in S & E problem-solving. In the rest of
the unstructured interviews (15), we also provided the experts with our list and asked them to comment
on any additions or deletions they would suggest. Faculty who had close supervision of graduate
students, and industry experts who had extensively supervised inexperienced staff, were particularly
informative. Their observations of the way inexperienced people could fail made them sensitive to the
different elements of expertise and where incorrect decisions could be made. Although we initially
expected to find substantial differences across disciplines, from early in the process we noted a high
degree of overlap across the interviews in the decisions that were described.

**Table 1: Interviews conducted, by field and position of interviewee.** "URM" included 3 African
American and 2 Hispanic.

| Discipline | Position of interviewee | | Informal interviews (creation of initial list) | Structured interviews (validation/ refinement) | Notes |
|---|---|---|---|---|---|
| Biology (5 biochemistry or molecular biology, 2 cell biology, 1 immunology, 1 plant biology, 1 ecology) | Total | | 2 | 8 | 6 female 2 URM |
| | • Academic faculty | | 2 | 3 | |
| | • Industry | | - | 2 | 1 also faculty, but counted here |
| | • Academic staff or postdoc (year 5+) | | - | 3 | |
| Medicine (6 internal medicine or pediatrics, 1 oncology, 2 surgery) | Total (All medical faculty) | | 4 | 6 | 4 female 1 URM (10 interviews with 9 faculty – one in both informal & structured) |

| Physics (4 experimental, 3 theoretical) | Total (All academic faculty) | 2 | | 5 | | 1 female 1 URM 2 physics +engineer |
|---|---|---|---|---|---|---|
| Electrical Engineering | Total | 4 | | 3 | | |
| | • Academic faculty | | 2 | - | | |
| | • Industry | | 1 | 3 | | |
| | • Academic staff | | 1 | - | | |
| Chemical Engineering | Total | 2 | | 2 | | 1 female |
| | • Industry | | 1 | 2 | | |
| | • Academic staff | | 1 | - | | |
| Mechanical Engineering | Total | 2 | | 2 | | 1 URM |
| | • Academic faculty | | 1 | 1 | | |
| | • Industry | | 1 | 1 | | |
| Earth Science | Total | 1 | | 2 | | 2 female |
| | • Academic faculty | | 1 | 1 | | |
| | • Industry | | - | 1 | | |
| Chemistry | Total (All academic faculty) | 1 | | 2 | | 2 female |
| Computer Science | Total | 2 | | 1 | | 1 female |
| | • Academic faculty | | 2 | - | | |
| | • Industry | | - | 1 | | |
| Biological Engineering | Total (All academic faculty or staff) | 2 | | - | | |
| Total | 53 interviews, with 52 experts | 22 | | 31 | | 17 female 5 URM |

**Refinement and validation of decisions list**

To refine and validate the list of decisions generated from the informal interviews, we then conducted semi-structured interviews with 31 experts from across science, engineering, and medical fields (Table 1). For these interviews, we recruited experts from a distribution of universities (including prestigious universities, state universities, and a liberal arts college) and companies. Interviews were conducted in person or over video chat and were transcribed for analysis. In the semi-structured interviews, experts were asked to choose a problem or two from their work that they could recall the details of solving, and then describe the process, including all the steps and decisions they made. So that we could get a full picture of the successful problem-solving process, we decided to focus the interviews on problems that they had eventually solved successfully, though their process inherently involved paths that needed to be revised and reconsidered.

Our interview protocol (see supplement) was inspired in part by the critical decision method of cognitive task analysis (Lintern et al., 2018; Crandall et al., 2006), which was created for research in cognitive systems engineering and naturalistic decision making. There are some notable differences between our work and theirs, both in research goal and method. First, their goal is to improve training in specific fields by focusing on *how* critical decisions are made in that field during an unusual or important event; the analysis seeks to identify factors involved in making those critical decisions. We are focusing on the overall problem-solving and how it compares across many different fields, which quickly led to attention on *what* are the decisions-to-be made, rather than how a limited set of those decisions are made. We asked experts to describe a specific, but not necessarily unusual, problem in their work, and focused our analysis on identifying all decisions made, not reasons for making them or identifying which were most critical. The specific order of problem-solving steps was also less important to us, in part because it was clear that there was no consistent order that was followed. Second, we are looking at different types of work. Cognitive systems engineering work has primarily focused on performance in professions like firefighters, power plant operators, military technicians, and nurses. These tend to require time sensitive critical skills that are taught with modest amounts of formal training. We are studying scientists, engineers, and doctors solving problems that require much longer and less time-critical solutions, and for which the formal training occupies many years.

Given our different focus, we made several adaptations to eliminate some of the more time-consuming steps from the interview protocol, allowing us to limit the interview time to approximately 1hr. Both protocols seek to elicit an accurate and complete reporting of the steps taken and decisions made in the process of solving a problem. Our general strategy was: 1) Have the expert explain their problem and talk step-by-step through the decisions involved in solving it, with relatively few interruptions from the interviewer except to keep the discussion focused on the specific problem and occasionally to ask for clarifications; 2) Ask follow-up questions to probe for more detail about particular steps and aspects of the problem-solving process; and 3) Occasionally ask for their general thoughts on how a novice's process might differ from theirs.

While some have questioned the reliability of information from retrospective interviews (Nisbett & Wilson, 1977), we believe we avoid these concerns, because we are only identifying a decision-to-be-made, which in this case, means identifying a well-defined action that was chosen from alternatives. This is less subjective and much more likely to be accurately recalled than is the rationale behind such a decision. See Ericsson and Simon (1980). We also are able to check the accuracy and completeness of the decisions identified by comparing them with the actions taken in the conduct of the research/design. For example, consider this quote from a physician who has to reevaluate a diagnosis, "*And, in my very subjective sense, he seemed like he was being forthcoming and honest. Granted people can fool you, but he seemed like he was being forthcoming. So we had to reevaluate*." They then considered alternative diagnoses that could explain a test result that, on first look, had indicated an incorrect diagnosis. While this quote does describe their (retrospective) reasoning behind their decision, we do not need to know whether that reasoning is accurately recalled. We can simply code this as "decisions #18 – info believable?" They follow up by considering alternative diagnoses, which in this context was coded as "#26 – solution good?" and "#8 – potential solutions," This was followed by the description of the literature and additional tests they conducted. These indicated actions taken that confirm they made a decision about the reliability of the information given by the patient.

**Interview coding**

We coded the semi-structured interviews in terms of decisions made, through iterative rounds of coding (Chi, 1997), following a "directed content analysis approach," which involves coding according to pre-defined theoretical categories and updating the codes as needed based on the data (Hsieh & Shannon, 2005). Our theoretical categories were the list of decisions we had developed during the informal interviews. The goals of each iterative round of coding are described below. To code for decisions in general, we matched decisions from the list to statements in each interview, based on the following criteria: 1) there was an explicit statement of a decision or choice made or needing to be made, 2) there was the description of the outcome of a decision, such as listing important features of the problem (that had been decided on) or conclusions arrived at, or 3) statement of actions taken that indicated a decision about the appropriate action had been made, usually from a set of alternatives. Two examples illustrate the types of comments we identified as decisions: A molecular biologist explicitly stated the decisions they needed to make in decomposing the problem into sub-problems (decision #11), "*Which cell do we use? The gene. Which gene do we edit? Which part of that gene do we edit? How do we build the enzyme that is going to do the cutting? … And how do we read out that it worked?*" An ecologist made a statement that was also coded as a decomposition decision, because they described the action taken: "*So I analyze the bird data first on its own, rather than trying to smash all the taxonomic groups together because they seem really apples and oranges. And just did two kinds of analysis, one was just sort of across all of these cases, around the world.*" A single statement could be coded as multiple decisions if they were occurring simultaneously in the story being recalled, or if they were intimately interconnected in the context of that interview, as with the ecology quote above where the last sentence leads into deciding what data analysis is needed. Inherent in nearly every one of these decisions was that there was insufficient information to know the answer with certainty, so judgement was required. Our primary goal for the first iterative round of coding was to check if our list was complete by checking for any decisions that were missing, as indicated by either an action taken or a stated decision that was not clearly connected to a decision on our initial list. In this round, we also clarified wording and combined decisions that we were consistently unable to differentiate during the coding. A sample of three interviews (from biology, medicine, and electrical engineering) were first coded independently by four coders, then discussed. The decision list was modified to add decisions and update wording based on that discussion. Then the interviews were re-coded with the new list and re-discussed, leading to more refinements to the list. Two additional interviews (from physics and chemical engineering) were then coded by three coders and further similar refinements were made. Throughout the subsequent rounds of coding, we continued to check for missing decisions, but after the additions and adjustments made based on these five interviews, we did not identify any more missing decisions.

In our next round of coding, we focused on condensing overlapping decisions and refining wording to improve the clarity of descriptions as they applied across different disciplinary contexts and to ensure consistent interpretation by different coders. Two or three coders independently coded an additional 11 interviews, iteratively refining wording, condensing the list of decisions, and then using the updated list to code subsequent interviews. We condensed the list by combining decisions that either represented the same cognitive process taking place at different times, that were discipline-specific variations on the same decision, or that were sub-steps involved in making a larger decision. We noticed that some decisions were frequently co-coded with others, particularly in some disciplines. But if they were identified as distinct a reasonable fraction of the time in any discipline, we listed them as separate. This provided us with a list, condensed from 42 to 29 discrete decisions (plus 5 additional non-decision themes), that gave good consistency between coders.

Finally, we used the resulting codes to tabulate which decisions occurred in each interview, simplifying our coding process to focus on deciding whether or not each decision had occurred, with an example if it did occur to back up the "yes" code, but no longer attempting to capture every time each decision was mentioned. Individual coders identified decisions mentioned in the remaining 15 interviews. Interviews that had been coded with the early versions of the list were also re-coded to ensure consistency. Coders flagged any decisions they were unsure about occurring in a particular interview, and 2-4 coders met to discuss those unsure codes, with most uncertainties being resolved by explanations from a team member who had more technical expertise in the field of the interview. Minor wording changes were made during this process to ensure that each description of a decision captured all instantiations of the decision across disciplines, but no significant changes to the list were needed or made.

Coding an interview in terms of decisions made and actions-taken-in-the-research often required a high level of expertise in the discipline in question. The coder had to be familiar with the conduct of research in the field in order to recognize which actions corresponded to a decision between alternatives, but our team was assembled with this requirement in mind. It included high level expertise across five different fields of science, engineering, and medicine, and substantial familiarity with several other fields.

Table S1 shows the final tabulation of decisions identified in each interview. In the tabulation, most decisions were marked as either "yes" or "no" for each interview, though 65 out of 1054 total were marked as "implied," for one of the following reasons: a) for 40/65, based on the coder's knowledge of the field, it was clear that a step must have been taken to achieve an outcome or action, even though that decision wasn't explicitly mentioned (for example, they describe collecting certain raw data and then coming to a specific conclusion, so they must have decided how to analyze the data, even if they didn't mention the analysis explicitly); b) for 15/65 the interview context was important, in that multiple statements from different parts of the interview taken together were sufficient to conclude that the decision must have happened, though no single statement described that decision explicitly; c) 10/65 involved a decision that was explicitly discussed as an important step in problem-solving but they did not directly state how it was related to the problem at hand, or it was stated only in response to a direct prompt from the interviewer. The proportion of decisions identified in each interview, broken down by either explicit or explicit + implied, is presented in Tables S1 and S2. Table 2 and Fig 2 of the main text show explicit + implied decision numbers.

To check inter-rater reliability using the final decisions list, two of the interviews that had not been discussed during earlier rounds of coding (one physics, one medicine) were independently coded by two coders. Which decisions were identified in each interview was compared for the two coders. For both interviews, they disagreed on whether or not only one of the 29 decisions occurred. Codes of "implied" were counted as agreement if the other coder selected either "yes" or "implied." This equates to a percent agreement of 97% for each interview, or a Cohen's Kappa of 0.72 for the physics interview and 0.84 for the medicine (Cohen's kappa is biased low if the prevalence of one code is high, as in these interviews when nearly all decisions are "yes" (Hallgren, 2012). As a side note, there was also one disagreement per interview on the coding of the 5 other themes, but those themes were not a focus of this work nor the interviews.

**Results**

We identified a total set of 29 decisions-to-be-made (plus 5 other themes), all of which were identified in a large fraction of the interviews across all disciplines (Table 2, Fig. 2). There was a surprising degree of overlap across the different fields with all the experts mentioning similar decisions-to-be-made. All 29 were evident by the fifth semi-structured interview, and on average, each interview revealed 85% of the twenty-nine decisions. Many decisions occurred multiple times in an interview, with the number of times varying widely, depending on the length and complexity of the solving process discussed.

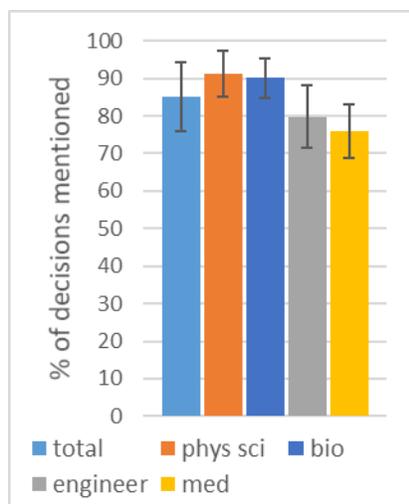

**Figure 2. Proportion of decisions coded in interviews by field.** This tabulation includes decisions #1-29, not the additional themes. Error bars represent standard deviations. Number of interviews: total 31, physical science 9, biological science 8, engineering 8, medicine 6. Compared to the sciences, slightly fewer decisions overall were identified in the coding of engineering and medicine interviews largely for discipline-specific reasons. See Table S4 and associated discussion.

**Table 2.** Problem-solving decisions [a] and percentages of expert interviews in which they occur.

| A. Selection and goals Occur in 100% [b] | 1. [c] What is important in field? 61% | 2. Opportunity fits solver's expertise? 77% | 3. Goals, criteria, constraints? 100% | |
|---|---|---|---|---|
| **B. Frame problem** 100% | 4. Important features and info? 100% | 5. What predictive framework? [d] 100% | 6. Narrow down problem 97% | 7. Related problems? 97% |
| | 8. Potential Solutions? 100% | 9. Is problem solvable? 74% | | |
| **C. Plan process for solving** 100% | 10. Approximations and simplifications 81% | 11. Decompose into sub-problems 68% | 12. Most difficult or uncertain areas? 90% | 13. What info needed? 100% |
| | 14. Priorities 87% | 15. Specific plan for getting information 100% | | |
| **D. Interpret information and choose solutions** 100% | 16. Calculations and data analysis 81% | 17. Represent and organize info 68% | 18. How believable is information? 77% | 19. Compare info to predictions 100% |
| | 20. Any significant anomalies? 71% | 21. Appropriate conclusions? 97% | 22. What is best solution? 97% | |
| **E. Reflect** [e] 100% | 23. Assumptions + simplifications appropriate? 77% | 24. Additional knowledge needed? 84% | 25. How well is solving approach working? 94% | 26. How good is solution? 100% |
| **F. Implications and communication of results** 84% | 27. Broader implications? 65% | 28. Audience for communication? 55% | 29. Best way to present work? 68% | |
| **G. Knowledge and skill dev.** [f] 100% | Stay up to date in field 84% | Intuition and experience 77% | Interpersonal, teamwork 100% | Efficiency 32% | Attitude 68% |

Footnotes: a) See supplementary text and Table S2 for full description and examples of each decision. b) Percentage of interviews in which category or decision was mentioned. c) Numbering is for reference. In practice ordering is fluid – involves extensive iteration with other possible starting points. d) Chosen predictive framework(s) will inform all other decisions. e) Reflection occurs throughout process, and often leads to iteration. Reflection on solution occurs at the end as well. f) Not decisions – other themes mentioned frequently as important to professional success.

We focused our analysis on what decisions needed to be made, not on the experts' process for making that decision: noting that a choice happened, not *how* they selected and chose among different alternatives. This is because, while the decisions-to-be-made were the same across disciplines, how the experts made those decisions varied greatly by discipline and individual. The process of making the decisions relied on specialized disciplinary knowledge and experience and may vary depending on demographics or other factors that our study design (both our sample and nature of retrospective interviews) did not allow us to investigate. However, while that knowledge was distinct and specialized, we could tell that it was consistently organized according to a common structure we call a "predictive framework," as discussed below. Also, while every "decision" reflected a step in the problem solving involved in the work, and the expert being interviewed was involved in making or approving the decisions, that does not mean the decision process was carried out only by that individual. In many cases, the experts described the decisions made in terms of ideas and results of their team, and the importance of interpersonal skills and teamwork was an important non-decision theme raised in all interviews.

We were particularly concerned with the correctness and completeness of the set of decisions. Although the correctness was largely established by the statements in the interviews, we also showed the list of decisions to these experts at the end of the interviews as well as to about a dozen other experts. In all cases, they all agreed that these decisions were ones they and others in their field made when solving problems. The completeness of the list of decisions was confirmed by: 1) we looked carefully at all specific actions taken in the described problem-solving and checked that each action matched with making a corresponding decision from the list; and 2) there was a high degree of consistency in the set of decisions across all the interviews and disciplines. This implies that it is unlikely that there are important decisions that we are missing, because that would require any such missing decisions to be consistently unspoken by all 31 interviewees as well as consistently unrecognized by us from the actions that were taken in the problem solving.

In focusing on experts' recollections of their *successful* solving of problems, our study design may have missed decisions that experts only made during failed problem-solving attempts. However, almost all interviews described solution paths that were not smooth and continuous, but rather involved going down numerous dead ends. There were approaches that were tried and failed, data that turned out to be ambiguous and worthless, etc. Identifying the failed path involved reflection decisions (#23-26). Often decision #9 (problem solvable?) would be mentioned, because they described a path that they decided was not solvable. For example, a biologist explained, "*And then I ended up just switching to a different strain that did it [crawling off the plate] less. Because it was just… hard to really get them to behave themselves. I suppose if I really needed to rely on that very particular one, I probably would have exhausted the possibilities a bit more.*" Thus, we expect unsuccessful problem-solving would entail a smaller subset of decisions being made, particularly lack of reflection decisions, or poor choices on the decisions, rather than making a different set of decisions.

The set of decisions represent a remarkably consistent structure underlying S & E problem-solving. For the purposes of presentation, we have categorized the decisions as shown in Fig. 3, roughly based on the purposes they achieve. However, the process is far less orderly and sequential than implied by this diagram, or in fact any characterization of an orderly "scientific method." We were struck by how variable the sequence of decisions was in the descriptions provided. For example, experts who described how they began work on a problem sometimes discussed importance and goals (#1-3 – importance, fit, goals), but others mentioned a curious observation (#20 - anomalies?), important features of their system that led them to questions (#4 – features, #6 - narrow), or other starting points. We also saw that there were flexible connections between decisions and repeated iterations – jumping back to the same type of decision multiple times in the solution process, often prompted by reflection as new information and insights were developed. The sequence and number of iterations described varied dramatically by interview, and we cannot determine to what extent this was due to legitimate differences in problem-solving process or to how the expert recalled and chose to describe their process. This lack of a consistent starting point, with jumping and iterating between decisions, has also been identified in the naturalistic decision making literature (Mosier et al., 2018). Finally, the experts also often described considering multiple decisions simultaneously. In some interviews, a few decisions were always described together, while in others they were clearly separate decisions. In summary, while the specific decisions themselves are fully grounded in expert practice, the categories and order shown here are artificial simplifications for presentation purposes.

The decisions contained in the seven categories are summarized below. See Table S2 for specific examples of each decision across multiple disciplines.

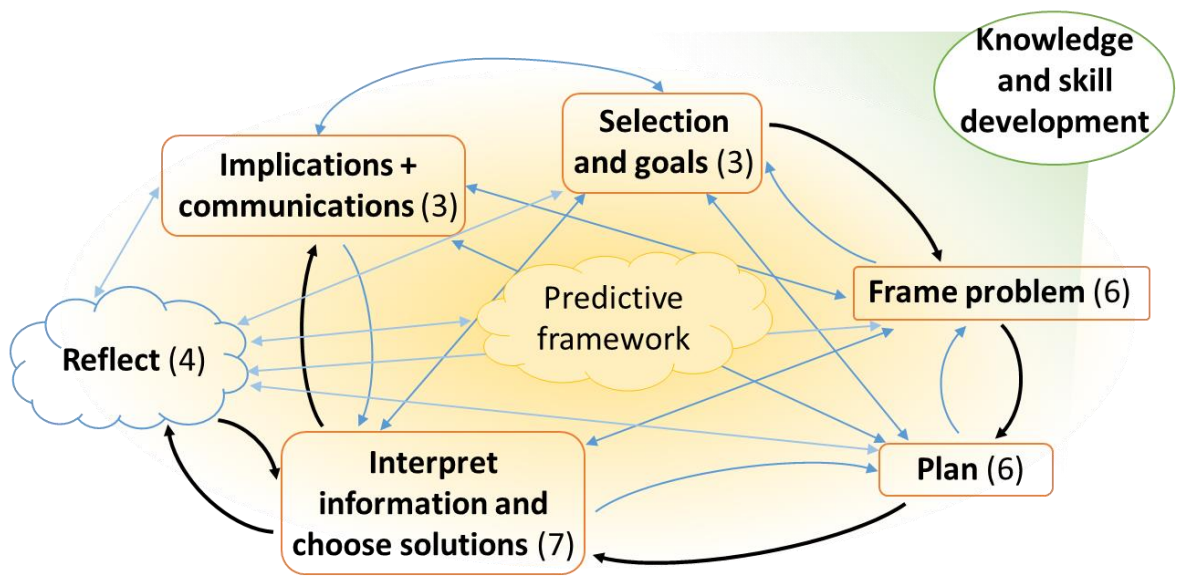

**Figure 3. Representation of problem-solving decisions by categories.** The black arrows represent a hypothetical but unrealistic order of operations, the blue arrows represent more realistic iteration paths. Numbers indicate the number of decisions in the category. Knowledge and skill development were commonly mentioned themes but are not decisions.

The decisions contained in the seven categories are summarized below. See Table S4 for specific examples of each decision across multiple disciplines.

**A. Selection and goals of the problem**

This category involves deciding on the importance of the problem, what criteria a solution must meet, and how well it matches the capabilities, resources, and priorities of the expert. As an example, an earth scientist described the goal of her project (decision #3 - goals) to map and date the earliest volcanic rocks associated with what is now Yellowstone, explained why the project was a good fit for her group (#2 - fit), and explained her decision to pursue the project in light of the significance of this type of eruption in major extinction events (#1 - importance). In many cases, decisions related to framing (see B below) were mentioned before decisions in this category or were an integral part of the process for developing goals.

Decisions in this category are:

1) **What is important in field?**

   What are important questions or problems? Where is the field heading? Are there advances in the field that open new possibilities?

2) **Opportunity fits solver's expertise?**

   If and where are there gaps/opportunities to solve in field? Given experts' unique perspectives and capabilities, are there opportunities particularly accessible to them? (could involve challenging the status quo, questioning assumptions in the field)

3) **Goals, criteria, constraints?**

   What are the goals for this problem? Possible considerations include:

   a. What are the goals, design criteria, or requirements of the problem or its solution?

   b. What is the scope of the problem?

   c. What constraints are there on the solution?

   d. What will be the criteria on which the solution is evaluated?

**B. Frame problem**

These decisions lead to a more concrete formulation of the solution process and potential solutions. This involves identifying the key features of the problem and deciding on predictive frameworks to use, as discussed below, as well as narrowing down the problem, often forming specific questions or hypotheses. Many of these decisions are guided by past problem solutions with which the expert is familiar and sees as relevant. The framing decisions of a physician can be seen in his discussion of a patient with liver failure who had previously been diagnosed with HIV, but had features (#4 - features, #5 – what predictive framework) that made the physician question the HIV diagnosis (#5 – what predictive framework, #26 – solution good?). His team then searched for possible diagnoses that could explain liver failure and lead to a false-positive HIV test (#7 – related problems, #8 – potential solutions), which led to their hypothesis the patient might have Q fever (#6 – narrow problem, #13 – info needed, #15 - plan). While each individual decision is strongly supported by the data, the categories are groupings for presentation purposes. In particular, framing (B) and planning (see C below) decisions often blended together in interviews.

Decisions in this category are:

4) **Important features and info?**

   What are the important underlying features or concepts that apply? Could include:

   a. Which available information is relevant to solving and why?

   b. (when appropriate) Create/find a suitable abstract representation of core ideas and information Examples: physics – equation representing process involved, chemistry – bond diagrams/potential energy surfaces, biology – diagram of pathway steps.

5) **What predictive framework?**

Which potential predictive frameworks to use? (decide among possible predictive frameworks or create framework) This includes deciding on the appropriate level of mechanism and structure that the framework needs to embody to be most useful for the problem at hand.

6) **Narrow down problem.**

How to narrow down the problem? Often involves formulating specific questions and hypotheses.

7) **Related problems?**

What are related problems or work seen before, and what aspects of their solving process and solutions might be useful in the present context? (may involve reviewing literature and/or reflecting on experience)

8) **Potential solutions?**

What are potential solutions? (Based on experience and fitting some criteria for solution they have for a problem having general key features identified.)

9) **Is problem solvable?**

Is the problem plausibly solvable and is solution worth pursuing given the difficulties, constraints, risks, and uncertainties?

## C. Plan process for solving

These decisions establish the specifics needed to solve the problem and include: how to simplify the problem and decompose it into pieces, what specific information is needed, how to obtain that information, and what are the resources needed and priorities? Planning by an ecologist can be seen in her extensive discussion of her process of simplifying (#10 – approximations/simplifications) a meta-analysis project about changes in migration behavior, which included deciding what types of data she needed (#13 – info needed), planning how to conduct her literature search (#15 - plan), difficulties in analyzing the data (#12 – difficult/uncertain areas?, #16 – data analysis), and deciding to analyze different taxonomic groups separately (#11 - decompose). In general, decomposition often resulted in multiple iterations through the problem-solving decisions, as sub-sets of decisions need to be made about each decomposed aspect of a problem. Framing (B) and planning (C) decisions occupied much of the interviews, indicating their importance.

Decisions in this category are:

10) **Approximations and simplifications.**

What approximations or simplifications are appropriate? How to simplify the problem to make it easier to solve? Test possible simplifications/approximations against established criteria.

11) **Decompose into sub-problems.**

How to decompose the problem into more tractable sub-problems? (Independently solvable pieces with their own sub-goals.)

12) **Most difficult or uncertain areas?**

Which are areas of particular difficulty and/or uncertainty in plan of solving process? Could include deciding:

   a.   What are acceptable levels of uncertainty with which to proceed at various stages?

13) **What info needed?**

What information is needed to solve the problem? Could include:

   a.   What will be sufficient to test and distinguish between potential solutions?

14) **Priorities.**

What to prioritize among many competing considerations? What to do first and how to obtain necessary resources?

Considerations could include: What's most important? Most difficult? Addressing uncertainties? Easiest? Constraints (time, materials, etc.)? Cost? Optimization and trade-offs? Availability of resources? (facilities/materials, funding sources, personnel)

15) **Specific plan for getting information**.

What is the specific plan for getting additional information? Includes:

a. What are the general requirements of a problem-solving approach, and what general approach will they pursue? (often decided early in problem-solving process as part of framing)

b. How to obtain needed information? Then carry out those plans (b.2). (This could involve many discipline and problem-specific investigation possibilities such as: designing and conducting experiments, making observations, talking to experts, consulting the literature, doing calculations, building models, or using simulations.)

c. What are achievable milestones, and what are metrics for evaluating progress?

d. What are possible alternative outcomes and paths that may arise during problem-solving process, both consistent with predictive framework and not, and what would be paths to follow for the different outcomes?

**D. Interpret information and choose solution(s)**

This category includes deciding how to analyze, organize, and draw conclusions from available information, reacting to unexpected information, and deciding upon a solution. A biologist studying aging in worms described how she analyzed results from her experiments, which included representing her results in survival curves and conducting statistical analyses (#16 – data analysis, #17 – represent and organize), as well as setting up blind experiments (#15 - plan) so that she could make unbiased interpretations (#18 – info believable?) of whether a worm was alive or dead. She also described comparing results to predictions to justify the conclusion that worm aging was related to fertility (#19 – compare to predictions, #21 – appropriate conclusions?, #22 – choose solution). Deciding how results compared with expectations based on a predictive framework was a key decision that often preceded several other decisions.

Decisions in this category are:

16) **Calculations and data analysis.**

What calculations and data analysis are needed? Then to carry those out.

17) **Represent and organize information.**

What is the best way to represent and organize available information to provide clarity and insights? (usually this will involve specialized & technical representations related to key features of predictive framework)

18) **How believable is information?**

Is information valid, reliable, and believable (includes recognizing potential biases)?

19) **Compare to predictions.**

As new information comes in, particularly from experiments or calculations, how does it compare with expected results (based on their predictive framework)?

20) **Any significant anomalies?**

If a result is different than expected, how should they follow up? (Requires first noticing the potential anomaly). Could involve deciding:

a. Does potential anomaly fit within acceptable range of predictive framework(s) (given limitations of predictive framework and underlying assumptions and approximations)?

b. Is potential anomaly an unusual statistical variation, or relevant data? Is it within acceptable levels of uncertainty?

21) **Appropriate conclusions?**

What are appropriate conclusions based on the data? (involves making conclusions and deciding if they're justified)

22) **What is the best solution?**

Deciding on best solution(s) involves evaluating and refining candidate solutions throughout problem-solving process. Not always narrowed down to a single solution. May include deciding:

a. Which of multiple candidate solutions are consistent with all available information and which can be rejected? (could be based on comparing data with predicted results)

b. What refinements need to be made to candidate solutions?

**E. Reflect**

Reflection decisions occur throughout the process and include deciding whether assumptions are justified, whether additional knowledge or information is needed, how well the solution approach is working, and if potential and then final solutions are adequate. These decisions match the categories of reflection identified by Salehi (2018). A mechanical engineer described developing a model (to inform surgical decisions) of which muscles allow the thumb to function in the most useful manner (#22 – choose solution), including reflecting on how well engineering approximations applied in the biological context (#23 – assumptions appropriate?). He also described reflecting on his approach – on why he chose to use cadaveric models instead of mathematical models (#25 – approach working?), and the limitations of his findings in that the "best" muscle they identified was difficult to access surgically (#26 – solution good?, #27 – implications). Reflection decisions are made throughout the problem-solving process, often lead to reconsidering other decisions, and are critical for success.

Decisions in this category are:

23) **Assumptions + simplifications appropriate?**

Are previous decisions about simplifications and predictive frameworks still appropriate?

a. Do the assumptions and simplifications made previously still look appropriate considering new information? (reflect on assumptions)

b. Does predictive framework need to be modified? (Reflect on predictive framework.)

24) **Additional knowledge needed?**

Is additional knowledge/information needed? (Based on ongoing review of one's state of knowledge.) Could involve:

a. Is solver's relevant knowledge sufficient?

b. Is more information needed and if so, what?

c. Does some information need to be checked? (e.g. need to repeat experiment or check a different source?)

25) **How well is solving approach working?**

How well is the problem-solving approach working, and does it need to be modified, including do the goals need to be modified? (Reflect on strategy by evaluating progress toward solution)

26) **How good is solution?**

How adequate is the chosen solution? (Reflect on solution) Includes ongoing reflection on potential solutions, as well as final reflection after selecting preferred solution. Can include:

     a.   Decide by exploring possible failure modes and limitations – "try to break" solution.

     b.   Does it "make sense" and pass discipline-specific tests for solutions of this type of problem?

     c.   Does it completely meet the goals/criteria?

**F. Implications and communication**

These are decisions about the broader implications of the work, and how to communicate results most effectively. For example, a theoretical physicist developing a method to calculate the magnetic moment of the muon decided on who would be interested in his work (#28 - audience) and what would be the best way to present it (#29 - present). He also discussed the implications of preliminary work on a simplified aspect of the problem (#10 – approximations/simplifications) in terms of evaluating its impact on the scientific community and deciding on next steps (#27 - implications, #29 - present). Many interviews described that making decisions in this category affected their decisions in other categories.

Decisions in this category are:

27) **Broader implications?**

What are the broader implications of the results, including over what range of contexts does the solution apply? What outstanding problems in field might it solve? What novel predictions can it enable? How and why might this be seen as interesting to a broader community?

28) **Audience for communication?**

What is the audience for communication of work, and what are their important characteristics?

29) **Best way to present work?**

What is the best way to present the work to have it understood, and its correctness and importance appreciated? How to make a compelling story of the work?

**G. Ongoing skill and knowledge development**

Although we focused on decisions in the problem-solving process, the experts volunteered general skills and knowledge they saw as important elements of problem-solving expertise in their field. These included teamwork and interpersonal skills (strongly emphasized), acquiring experience and intuition, and keeping abreast of new developments in their field.

Non-decision themes in this category are:

(30) **Stay up to date in field**

Staying up to date could include:

a. Review literature, which does involve making decisions as to which is important.

b. Learn relevant new knowledge (ideas and technology, from literature, conferences, colleagues, etc.)

(31) **Intuition and experience.**

Acquiring experience and associated intuition to improve problem-solving.

(32) **Interpersonal, teamwork**.

Includes navigating collaborations, team management, patient interactions, communication skills, etc., particularly as how these apply in the context of the various types of problem-solving processes.

(33) **Efficiency.**

Time management including learning to complete certain common tasks efficiently and accurately.

(34) **Attitude.**

Motivation and attitude to the task. Factors such as interest, perseverance, dealing with stress, confidence in decisions, etc.

**Predictive framework**

How the decisions were made was highly dependent on the discipline and problem. However, there was one element that was fundamental and common across all interviews: the early adoption of a "predictive framework" that the experts used throughout their problem-solving process. We define this framework as "a mental model of key features of the problem and the relationships between the features." All the predictive frameworks involved some degree of simplification and approximation and an underlying level of mechanism that established the relationships between key features. The frameworks provided a structure of knowledge and facilitated the application of that knowledge to the problem at hand, allowing experts to repeatedly run "mental simulations" to make predictions for dependencies and observables and to interpret new information.

As an example, an ecologist described her predictive framework for migration, which incorporated important features such as environmental conditions and genetic differences between species and the mechanisms by which these interacted to impact the migration patterns for a species. She used this framework to guide her meta-analysis of changes in migration patterns, affecting everything from her choice of datasets to include to her interpretation of why migration patterns changed for different species. In many interviews, the frameworks used evolved as additional information was obtained, with additional features being added or underlying assumptions modified. For some problems, the relevant framework was well established and used with confidence, while for other problems there was considerable uncertainty as to a suitable framework, so developing and testing the framework was a substantial part of the solution process.

A predictive framework contains the expert knowledge organization that has been observed in previous studies of expertise (Egan & Greeno, 1974) but goes farther, as here it serves as an explicit tool that guides most decisions and actions during the solving of complex problems. Mental models and mental simulations that are described in the naturalistic decision making literature are similar, in that they are used to understand the problem and guide decisions (Klein, 2008; Mosier et al., 2018), but they do not necessarily contain the same level of mechanistic understanding of relationships that underlies the predictive frameworks used in science and engineering problem-solving. While the use of predictive frameworks was universal, the individual frameworks themselves explicitly reflected the relevant specialized knowledge, structure, and standards of the discipline, and arguably largely define a discipline (Wieman, 2019).

**Discipline-specific variation**

While the set of decisions-to-be-made was highly consistent across disciplines, there were extensive differences within and across disciplines and work contexts, which reflected the differences in perspectives and experiences. These differences were usually evident in how they made each of the specific decisions, but not in the choice of which decisions needed to be made. In other words, the solution methods, which included following standard accepted procedures in each field, were very different. For example, planning in some experimental sciences may involve formulating a multiyear construction and data collection effort, while in medicine it may be deciding on a simple blood test. Some decisions, notably in categories A, D, and F, were less likely to be mentioned in particular disciplines, because of the nature of the problems. Specifically, decisions #1 (importance), 2 (fit), 27 (implications), 28 (audience), and 29 (best way to present) were dependent on the scope of the problem being described and the expert's specific role in it. These were mentioned less frequently in interviews where the problem was assigned to the expert (most often engineering or industry) or where the importance or audience was implicit (most often in medicine). Decisions 16 (data analysis) and 17 (represent/organize) were particularly unlikely to be mentioned in medicine, because test results are typically provided to doctors not in the form or raw data, but rather already analyzed by a lab or other medical technology professional, so the doctors we interviewed did not need to make decisions themselves about how to analyze or represent the data. Qualitatively, we also noticed some differences between disciplines in the patterns of connections between decisions. When the problem involved development of a tool or product, most commonly the case in engineering, the interview indicated relatively rapid cycles between goals (#3), framing problem/potential solutions (#8), and reflection on potential solution (#26), before going through the other decisions. Biology, the experimental science most represented in our interviews, had strong links between planning (#15), deciding on appropriate conclusions (#21), and reflection on solution (#26). This is likely because the respective problems involved complex systems with many unknowns, so careful planning was unusually important for achieving definitive conclusions. See supplemental text and Table S2 for additional notes on decisions that were mentioned at lower frequency and decisions that were likely to be interconnected, regardless of field.

**Discussion**

This work has created a theoretical framework to characterize problem-solving in science and engineering. This framework is empirically based and captures the successful problem-solving process of all experts interviewed. We see that several dozen experts across many different fields all make a common set of decisions when solving authentic problems. There are flexible linkages between decisions that are guided by reflection in a continually evolving process. We have also identified the nature of the "predictive frameworks" that S & E experts consistently use in problem-solving. These predictive frameworks reveal how these experts organize their disciplinary knowledge to facilitate making these decisions. Many of the decisions we identified are reflected in previous work on expertise and scientific problem solving. This is particularly true for those listed in "planning" and "interpreting information" categories (Egan & Greeno, 1974). The priority experts give to framing and planning decisions over execution compared to novices has been noted repeatedly (e. g. Chi et al., 2015). Expert reflection has been discussed, but less extensively (Chase & Simon, 1973), and elements of the "selection" and "implications and communication" categories have been included in policy and standards reports (e.g. AAAS, 2011). Thus, our framework of decisions is consistent with previous work on "scientific practices" and expertise, but it is more complete, specific, empirically based, and generalizable across S & E disciplines.

A limitation of this study is the small number of experts we have in total and from each discipline, and the lack of randomized selection of participants. This means we cannot prove that there are not some experts who follow other paths in problem-solving. However, to our knowledge, this is a far larger sample than used in any previous study of expert problem-solving. Although we see a large amount of variation both within and across disciplines in the problem solving, this is reflected in how they make decisions, not in what decisions they make. The very high degree of consistency in the decisions made across the entire sample strongly suggests that we are capturing elements that are common to all experts across science and engineering. A second limitation is that decisions often overlap and co-occur in an interview, so the division between decision items is often somewhat ambiguous and could be defined somewhat differently. As noted, a number of these decisions can be interconnected, and in some fields are nearly always interconnected.

The set of decisions we have observed provides a general framework for characterizing, analyzing, and teaching S & E problem-solving. These decisions likely define much of the set of cognitive skills a student needs to practice and master to perform as a skilled practitioner in S & E. This framework of decisions provides a detailed and structured way to approach the teaching and measurement of problem-solving at the undergraduate, graduate, and professional training levels. For teaching, we propose using the process of "deliberate practice" (Ericsson, 2018) to help students learn problem solving. They would have more effective scaffolding and concentrated practice, with feedback, at making the specific decisions identified here in relevant contexts. In a course, this would likely involve only an appropriately selected set of the decisions, but a good research mentor would ensure their trainees have opportunities to practice and receive feedback on their performance on each of these 29 decisions. Measurements of individual problem-solving expertise based on our decision list and the associated discipline-specific predictive frameworks will allow a detailed measure of an individual's discipline-specific problem-solving strengths and weaknesses relative to an established expert. This can be used to provide targeted feedback to the learner, and when aggregated across students in a program, feedback on the educational quality of the program. We are currently working on the implementation of these ideas in a variety of instructional settings and will report on that work in future publications.

As discussed in the introduction, typical science and engineering problems fail to engage students in the complete problem-solving process. By considering which of the 29 decisions are required to answer the problem, we can more clearly articulate why. The biology problem, for example, requires students to decide on a predictive framework and access the necessary content knowledge, and they need to decide which information they need to answer the problem. However, other decisions are already made for them, such as deciding on important features and identifying anomalies, or not required. We propose that different problems, designed specifically to require students to make sets of the problem-solving decisions from our framework, will provide more effective tools for measuring, practicing, and ultimately mastering the full S & E problem-solving process.

Our preliminary work with the use of such decision-based problems for assessing problem-solving expertise is showing great promise. For several different disciplines, we have given test subjects a relevant context, requiring content knowledge covered in courses they have taken, and asked them to make decisions from the list presented here. Skilled practitioners in the relevant discipline respond in very consistent ways, while students respond very differently and show large differences which typically correlate with their different educational experiences. What apparently matters is not what content they have seen, but rather, what decisions they have had practice making. What we have seen is consistent with previous work identifying expert-novice differences but provides a much more extensive and detailed picture of a student's strengths and weaknesses and the impacts of particular educational experiences. We have also carried out preliminary development of courses that explicitly involve students making and justifying many of these decisions in relevant contexts, followed by feedback on their decisions. Preliminary results from these courses are also encouraging. Future work will involve the more extensive development and application of decision-based measurement and teaching of problem-solving.

**Acknowledgments**

We acknowledge the many experts who agreed to be interviewed for this work, M. Flynn for contributions on expertise in mechanical engineering, and Shima Salehi for useful discussions. This work was funded by the Howard Hughes Medical Institute through an HHMI Professor grant to C. Wieman.

**References:**

ABET. (2020). Criteria for Accrediting Engineering Programs, 2020 – 2021. Retrieved from

https://www.abet.org/accreditation/accreditation-criteria/criteria-for-accrediting-engineering-

programs-2020-2021/

Alberts, B., Johnson, A., Lewis, J., Morgan, D., Raff, M., Roberts, K., & Walter, P. (2014). Control of gene

expression. In *Molecular Biology of the Cell, 6th Edition* (pp. 436-437). Garland Science, Taylor &

Francis Group, LLC. Retrieved from https://books.google.com/books?id=2xIwDwAAQBAJ

American Association for the Advancement of Science (AAAS). (2011). Vision and Change in

Undergraduate Biology Education: A Call to Action, Washington, DC. Retrieved from

https://visionandchange.org/finalreport/

Chase, W.G., & Simon, H.A. (1973). Perception in chess. *Cognitive Psychology* 4(1), 55-81.

https://doi.org/10.1016/0010-0285(73)90004-2

Chi, M.T.H. (1997). Quantifying Qualitative Analyses of Verbal Data: A Practical Guide. *Journal of the

Learning Sciences* 6(3), 271–315. https://doi.org/10.1207/s15327809jls0603_1

Chi, M.T.H., Feltovich, P.J., & Glaser, R. (1981). Categorization and representation of physics problems by experts and novices. *Cognitive Science,* 5(2), 121-152.

https://doi.org/10.1207/s15516709cog0502_2

Chi, M.T.H., Glaser, R., & Farr, M.J., (Eds.). (2015). *The nature of expertise*. Lawrence Erlbaum Associates,

Inc.

Clemmons, A.W., Timbrook, J., Herron, J.C., & Crowe, A.J. (2020). BioSkills guide: Development and

national validation of a tool for interpreting the *vision and change* core competencies. *CBE-Life

Sciences Education,* 19(4), https://doi.org/10.1187/cbe.19-11-0259

Crandall, B., Klein, G.A., & Hoffman, R.R. (2006). *Working Minds: A Practitioner's Guide to Cognitive Task

Analysis* MIT Press.

Dunbar, K. (2000). How Scientists Think in the Real World: Implications for Science Education. *Journal of Applied Developmental Psychology* 21(1), 49–58. https://doi.org/10.1016/S0193-3973(99)00050-7

Egan, D.E., & Greeno, J.G. (1974). Theory of rule induction: Knowledge acquired in concept learning, serial pattern learning, and problem solving in L.W. Gregg (Ed.), *Knowledge and Cognition*. Lawrence Erlbaum.

Ericsson, K.A., & Simon, H.A. (1980). Verbal reports as data. *Psychological Review* 87(3), 215–251. https://doi.org/10.1037/0033-295X.87.3.215

Ericsson, K.A., Charness, N., Feltovich, P.J., & Hoffman, R.R., (Eds.). (2006). *The Cambridge handbook of expertise and expert performance*. Cambridge University Press.

Ericsson, K.A., Hoffman, R.R., Kozbelt, A., & Williams, A.A., (Eds.). (2018). *The Cambridge handbook of expertise and expert performance, 2nd ed.* Cambridge University Press.

Ericsson, K.A. (2018). The differential influence of experience, practice, and deliberate practice on the development of superior individual performance of experts. In K.A. Ericcson, R.R. Hoffman, A. Kozbelt, & A. M. Williams (Eds.), *The Cambridge handbook of expertise and expert performance* (pp. 745-769). Cambridge University Press. https://doi.org/10.1017/9781316480748.038

Greeno, J. G. (1994). Gibson's affordances. *Psychological Review* 101(2), 336-342. doi:10.1037/0033-295X.101.2.336

Hallgren, KA. (2012). Computing inter-rater reliability for observational data: An overview and tutorial. *Tutor Quant Methods Psychol*. 8(1): 23-34.

Hatano, G., & Inagaki, K. (1986). Two courses of expertise. In H. W. Stevenson, H. Azuma, & K. Hakuta (Eds.), *A series of books in psychology. Child development and education in Japan* (p. 262–272). W H Freeman/Times Books/ Henry Holt & Co.

Heckler, A.F. (2010). Some Consequences of Prompting Novice Physics Students to Construct Force

Diagrams. *International Journal of Science Education* 32(14), 1829–1851.

https://doi.org/10.1080/09500690903199556

Heller, J.I., & Reif, F. (1984). Prescribing Effective Human Problem-Solving Processes: Problem

Description in Physics. *Cognition and Instruction,* 1(2), 177–216.

https://doi.org/10.1207/s1532690xci0102_2

Heller, P., Keith, R., & Anderson, S. (1992). Teaching problem solving through cooperative grouping. Part

1: Group versus individual problem solving. *American Journal of Physics* 60, 627–636.

https://doi.org/10.1119/1.17117

Hsieh, H-F, & Shannon, S.E. (2005). Three Approaches to Qualitative Content Analysis. *Qual Health Res*

15(9), 1277–1288 (2005). https://doi.org/10.1177/1049732305276687

Huffman, D. (1997). Effect of explicit problem-solving instruction on high school students' problem-

solving performance and conceptual understanding of physics. *Journal of Research in Science*

*Teaching* 34(6), 551–570. https://doi.org/10.1002/(SICI)1098-2736(199708)34:6<551::AID-

TEA2>3.0.CO;2-M

Kind, P., & Osborne, J. (2016). Styles of scientific reasoning: A cultural rationale for science education?

*Science Education,* 10(1), 8-31. https://doi.org/10.1002/sce.21251

Klein, G. (2008). Naturalistic decision making. *Human factors*, *50*(3), 456-460.

Kozma, R., Chin, E., Russell, J., & Marx, N. (2000). The roles of representations and tools in the chemistry

laboratory and their implications for chemistry learning. Journal of the Learning Sciences, 9(2),

105-143.

Kuo, E., Hallinen, N.R., & Conlin, L.D. (2017). When procedures discourage insight: epistemological

consequences of prompting novice physics students to construct force diagrams. *International*

*Journal of Science Education* 39(7), 814–839. https://doi.org/10.1080/09500693.2017.1308037

Larkin J. & Reif F, (1979). Understanding and Teaching Problem-Solving in Physics, Eur. J. of Science

    Education, 1:2, 191-203. DOI: 10.1080/0140528790010208

Lintern G., Moon B., Klein G., and Hoffman R., (2018). Chap. 11, Eliciting and Representing the

    Knowledge of Experts, in Ericsson, K.A., Hoffman, R.R., Kozbelt, A., & Williams, A.A., (Eds.). *The*

    *Cambridge handbook of expertise and expert performance, 2nd ed*. Cambridge University Press.

Mosier, K., Fischer, U., Hoffman, R. R., & Klein, G. (2018). Chap. 25, Expert Professional Judgments and

    "Naturalistic Decision Making", in Ericsson, K.A., Hoffman, R.R., Kozbelt, A., & Williams, A.A.,

    (Eds.). *The Cambridge handbook of expertise and expert performance, 2nd ed*. Cambridge

    University Press.

National Research Council (NRC). (2012a). 5 Problem Solving, Spatial Thinking, and the Use of

    Representations in Science and Engineering. *Discipline-Based Education Research:*

    *Understanding and Improving Learning in Undergraduate Science and Engineering*, pp. 75–118.

    https://doi.org/10.17226/13362

National Research Council (NRC). (2012b). *A Framework for K-12 Science Education: Practices,*

    *Crosscutting Concepts, and Core Ideas*. National Academies Press.

Newell, A., & Simon, H.A. (1972). *Human problem solving*. Prentice-Hall.

NGSS Lead States. (2013). *Next Generation Science Standards: For States, By States.* Washington, DC:

    The National Academies Press.

Nisbett, R.E., & Wilson, T.D. (1977). Telling more than we can know: Verbal reports on mental processes.

    *Psychological Review* 84(3), 231–259. https://doi.org/10.1037/0033-295X.84.3.231

OECD. (2019). PISA 2018 Science Framework in *PISA 2018 Assessment and Analytical Framework,* OECD

    Publishing, Paris (pp. 97–117). https://doi.org/10.1787/f30da688-en.

Polya, G. (1945). *How to Solve It: A New Aspect of Mathematical Method*. Princeton University Press.

Priemer, B., Eilerts, K., Filler, A., Pinkwart, N., Rösken-Winter, B., Tiemann, R., & Upmeier Zu Belzen, A.

(2020). A framework to foster problem-solving in STEM and computing education. *Research in*

*Science & Technological Education,* 38(1), 105-130.

https://doi.org/10.1080/02635143.2019.1600490

Quacquarelli Symonds (QS). (2018). The Global Skills Gap in the 21st Century.

Salehi, S. (2018). *Improving problem-solving through reflection.* [Doctoral dissertation. Stanford

University.] Stanford Digital Repository. https://purl.stanford.edu/gc847wj5876

Schoenfeld, A.H. (1985). *Mathematical Problem Solving*. Elsevier.

Simon, H. (1973). The structure of ill structured problems. *Artificial Intelligence*, 4(3-4), 181-201.

https://doi.org/10.1016/0004-3702(73)90011-8

Singh, C. (2002). When physical intuition fails. *American Journal of Physics* 70, 1103–1109.

https://doi.org/10.1119/1.1512659

Wieman, C.E. (2015). Comparative Cognitive Task Analyses of Experimental Science and Instructional

Laboratory Courses. *The Physics Teacher* **53**, 349–351. https://doi.org/10.1119/1.4928349

Wieman, C.E. (2019). Expertise in University Teaching & the Implications for Teaching Effectiveness,

Evaluation & Training. *Daedalus* 148(4), 47–78. https://doi.org/10.1162/daed_a_01760

Wineburg, S. (1998). Reading Abraham Lincoln: An Expert/Expert Study in the Interpretation of Historical

Texts. *Cognitive Science* 22(3), 319–346. https://doi.org/10.1016/S0364-0213(99)80043-3

**Supplementary Information for**

A Detailed Characterization of the Expert Problem-Solving Process in Science and Engineering; Guidance for Teaching and Assessment

Authors: Argenta M. Price, Candice J. Kim, Eric W. Burkholder, Amy V. Fritz, and Carl E. Wieman

**This file includes:**

    Supplementary text: interview protocol, full decisions list, notes on trends in specific decisions
    Legends for Tables S1 and S2
    Legend for Dataset S1

**Other supplementary materials for this manuscript include the following:**

    Tables S1 and S2
    Dataset S1

**Supplementary Information Text**

**Supplementary Methods**

**Complete semi-structured interview protocol**

*Notes:*

*Inspired by the "Critical Decision Method" protocol of cognitive task analysis* (Crandall et al., 2006)

*Semi-structured interview with many questions optional, depending on course of interview. Questions in bold were prioritized, so were usually asked.*

*Aside from initial prompts and prompts to keep them on target or to provide enough detail, interviewers will say very little during the story telling part of the interview, then will ask elaboration and deepening questions to follow up.*

<u>Introduction</u>

We are interviewing you as part of a project to identify how experts think as they solve problems during their research/work. Our goal is to identify what students ought to be learning to do in order to improve education. So today we'd like to learn how you solve problems by having you recall a problem you have solved or project you've completed, and walk us through all the detailed steps. Particularly focus on the detailed decisions you made when solving the problem.

<u>Eliciting the scenario/problem:</u>

1. **So, think about a problem you've solved in your work (or project you've completed). Choose a problem in which you can remember all the detailed steps and decisions.**

2. Then please walk me/us through how you went about solving that problem: What were the goals you were trying to achieve? What did you do step by step? What decisions did you make?

<u>Optional guidance questions to ask during story</u>

   a. If they're having trouble thinking of a problem:
      i. What is the most recent paper you've published/project you've worked on? How did you go about tackling that project?
      ii. What was a particularly challenging research problem (or design problem or work problem) you've dealt with?
   b. If they need more guidance starting to tell the story:
      i. What was the first decision you needed to make?
      ii. **What did you do first?**
         1. What did you do next? (phrase appropriately to respond to something in their narrative if they've stalled)
      iii. You mentioned X goal, what did you do to accomplish that?
      iv. What were the most important things for you to think about?

    c. If they give too short or high-level of an account
        i. (Especially at beginning to set the tone) probe for decisions made that were unstated: How did you decide X, that you just mentioned?
        ii. What led to your decision X?
        iii. What did you mean by X?

Check-in, clarifications, and elaboration (may be asked during story, as needed):

3. Ask for clarification about parts of the story that were unclear
    a. In particular, if they used specific words like "model," ask them to elaborate on their meaning of the term.
        i. Ask for examples of where and how they used term/models (if not already stated) – elaborating on meaning may be best done through examples.
4. Ask for elaboration on parts of the story that were glossed over (can interrupt story to ask, but give them time to get there themselves)
    a. How did you decide…?
    b. What led to your decision X?
    c. Why did (or didn't) you…?
    d. What did you do next?

More specific elaboration questions

*Note: Can ask after story telling if they weren't naturally covered, or don't need to ask if they came up on their own. Can also bring up during story to help move story along or away from excess detail or attempts to teach. Prioritize bolded questions (combine "elaboration" and "deepening").*

5. **How did you decide to tackle the problem in the first place?**
6. You said you did X first; how did you decide to tackle that aspect first?
    a. Or more general (preferred): How did you decide which way to go first?
7. You said you chose X method/route; how/why did you pick that method over other possible methods?
    a. Or more general (preferred): Why did you choose the path or methods you chose?
8. What information or data did you need to collect?
    a. Where did you get this information?
    b. What did you do with this information?
9. How did you interpret the results you collected (at X point)?
10. Were there other solutions you considered?
    a. How did you differentiate the possible solutions?

"Deepening" questions about the whole process

11. **What did you think were critical decisions you made during the process?**

      a. If they don't mention a particular point of story that interviewer thinks involved critical decisions, can ask for elaboration about decisions made during that particular point

      b. What decisions were you given vs. that you made (or what were the parameters you had at the start)?

**12. What challenges did you encounter in solving the problem?**

      a. How did you deal with those challenges?

**13. What (if any) new knowledge or skills did you need to acquire for solving your problem?**

      a. How did you acquire those?

14. What tests or experiments did you run?

      a. How did you decide on them?

      b. How did you interpret the results? (or ask alternative about a specific point – question 8)

15. (optional) How does this connect with prior work you've done?

**16. How did you decide you had an adequate result (to publish paper/submit design, etc.)**

      a. How did you know you were done?

      b. When did you decide your design/solution/conclusion was satisfactory?

17. What are the implications (or next steps) for your project?

18. How were your findings/product received by the community?

<u>What-if scenarios:</u>

19. What alternative decisions could you have made (in general or at specific point X), and what might have happened differently?

**20. If your student/trainee had been solving this problem instead of you (or without your guidance), how do you think their approaches would have differed from yours?**

<u>Other questions about perspective on expertise in their field (if time)</u>

**21. What are particular difficulties in problem-solving you've noticed in people you have trained or mentored? (be specific)**

**22. How do you use models in your work?**

23. What do you see as differences between experts and novices?

24. What do you think was your particular expertise that made you successful (in solving this problem, and more generally in your career)?

25. What do you want your trainees to be able to do (While trainees? After they're done training?)?

**Supplementary Text**

**Complete list of decisions-to-be made by experts when solving authentic problems**

    **A. Selection and goals**

      10) **What is important in field?**

What are important questions or problems? Where is the field heading? Are there advances in the field that open new possibilities?

11) **Opportunity fits solver's expertise?**

If and where are there gaps/opportunities to solve in field? Given experts' unique perspectives and capabilities, are there opportunities particularly accessible to them? (could involve challenging the status quo, questioning assumptions in the field)

12) **Goals, criteria, constraints?**

What are the goals for this problem? Possible considerations include:

    a. What are the goals, design criteria, or requirements of the problem or its solution?

    b. What is the scope of the problem?

    c. What constraints are there on the solution?

    d. What will be the criteria on which the solution is evaluated?

B. **Frame Problem**

13) **Important features and info?**

What are the important underlying features or concepts that apply? Could include:

    a. Which available information is relevant to solving and why?

    b. (when appropriate) Create/find a suitable abstract representation of core ideas and information Examples: physics – equation representing process involved, chemistry – bond diagrams/potential energy surfaces, biology – diagram of pathway steps.

14) **What predictive framework?**

Which potential predictive frameworks to use? (decide among possible predictive frameworks or create framework) This includes deciding on the appropriate level of mechanism and structure that the framework needs to embody to be most useful for the problem at hand.

15) **Narrow down problem.**

How to narrow down the problem? Often involves formulating specific questions and hypotheses.

16) **Related problems?**

What are related problems or work seen before, and what aspects of their solving process and solutions might be useful in the present context? (may involve reviewing literature and/or reflecting on experience)

17) **Potential solutions?**

What are potential solutions? (Based on experience and fitting some criteria for solution they have for a problem having general key features identified.)

18) **Is problem solvable?**

Is the problem plausibly solvable and is solution worth pursuing given the difficulties, constraints, risks, and uncertainties?

C. **Plan Process for Solving**

19) **Approximations and simplifications.**

What approximations or simplifications are appropriate? How to simplify the problem to make it easier to solve? Test possible simplifications/approximations against established criteria.

20) **Decompose into sub-problems.**

How to decompose the problem into more tractable sub-problems? (Independently solvable pieces with their own sub-goals.)

21) **Most difficult or uncertain areas?**

Which are areas of particular difficulty and/or uncertainty in plan of solving process? Could include deciding:

    a. What are acceptable levels of uncertainty with which to proceed at various stages?

22) **What info needed?**

What information is needed to solve the problem? Could include:

    a. What will be sufficient to test and distinguish between potential solutions?

23) **Priorities.**

What to prioritize among many competing considerations? What to do first and how to obtain necessary resources?

Considerations could include: What's most important? Most difficult? Addressing uncertainties? Easiest? Constraints (time, materials, etc.)? Cost? Optimization and trade-offs? Availability of resources? (facilities/materials, funding sources, personnel)

24) **Specific plan for getting information**.

What is the specific plan for getting additional information? Includes:

    a. What are the general requirements of a problem-solving approach, and what general approach will they pursue? (often decided early in problem-solving process as part of framing)

    b. How to obtain needed information? Then carry out those plans (b.2). (This could involve many discipline and problem-specific investigation possibilities such as: designing and conducting experiments, making observations, talking to experts, consulting the literature, doing calculations, building models, or using simulations.)

    c. What are achievable milestones, and what are metrics for evaluating progress?

    d. What are possible alternative outcomes and paths that may arise during problem-solving process, both consistent with predictive framework and not, and what would be paths to follow for the different outcomes?

**D. Interpret Information and Choose Solutions**

25) **Calculations and data analysis.**

What calculations and data analysis are needed? Then to carry those out.

26) **Represent and organize information.**

What is the best way to represent and organize available information to provide clarity and insights? (usually this will involve specialized & technical representations related to key features of predictive framework)

27) **How believable is information?**

Is information valid, reliable, and believable (includes recognizing potential biases)?

28) **Compare to predictions.**

As new information comes in, particularly from experiments or calculations, how does it compare with expected results (based on their predictive framework)?

29) **Any significant anomalies?**

If a result is different than expected, how should they follow up? (Requires first noticing the potential anomaly). Could involve deciding:

      a. Does potential anomaly fit within acceptable range of predictive framework(s) (given limitations of predictive framework and underlying assumptions and approximations)?

      b. Is potential anomaly an unusual statistical variation, or relevant data? Is it within acceptable levels of uncertainty?

30) **Appropriate conclusions?**

What are appropriate conclusions based on the data? (involves making conclusions and deciding if they're justified)

31) **What is the best solution?**

Deciding on best solution(s) involves evaluating and refining candidate solutions throughout problem-solving process. Not always narrowed down to a single solution. May include deciding:

      a. Which of multiple candidate solutions are consistent with all available information and which can be rejected? (could be based on comparing data with predicted results)

      b. What refinements need to be made to candidate solutions?

**E. Reflect (ongoing)**

32) **Assumptions + simplifications appropriate?**

Are previous decisions about simplifications and predictive frameworks still appropriate?

      a. Do the assumptions and simplifications made previously still look appropriate considering new information? (reflect on assumptions)

      b. Does predictive framework need to be modified? (Reflect on predictive framework.)

33) **Additional knowledge needed?**

Is additional knowledge/information needed? (Based on ongoing review of one's state of knowledge.) Could involve:

      a. Is solver's relevant knowledge sufficient?

      b. Is more information needed and if so, what?

      c. Does some information need to be checked? (e.g. need to repeat experiment or check a different source?)

34) **How well is solving approach working?**

How well is the problem-solving approach working, and does it need to be modified, including do the goals need to be modified? (Reflect on strategy by evaluating progress toward solution)

35) **How good is solution?** How adequate is the chosen solution? (Reflect on solution) Includes ongoing reflection on potential solutions, as well as final reflection after selecting preferred solution. Can include:

      a. Decide by exploring possible failure modes and limitations – "try to break" solution.

      b. Does it "make sense" and pass discipline-specific tests for solutions of this type of problem?

      c. Does it completely meet the goals/criteria?

**F. Implications and Communications of Results**

36) **Broader implications?**

What are the broader implications of the results, including over what range of contexts does the solution apply? What outstanding problems in field might it solve? What novel predictions can it enable? How and why might this be seen as interesting to a broader community?

37) **Audience for communication?**
What is the audience for communication of work, and what are their important characteristics?

38) **Best way to present work?**
What is the best way to present the work to have it understood, and its correctness and importance appreciated? How to make a compelling story of the work?

G. **Non-Decision Themes: Ongoing Knowledge and Skill Development**

(30) **Stay up to date in field**
Staying up to date could include:
   a. Review literature, which does involve making decisions as to which is important.
   b. Learn relevant new knowledge (ideas and technology, from literature, conferences, colleagues, etc.)

(31) **Intuition and experience.**
Acquiring experience and associated intuition to improve problem-solving.

(32) **Interpersonal, teamwork**.
Includes navigating collaborations, team management, patient interactions, communication skills, etc., particularly as how these apply in the context of the various types of problem-solving processes.

(33) **Efficiency.**
Time management including learning to complete certain common tasks efficiently and accurately.

(34) **Attitude.**
Motivation and attitude to the task. Factors such as interest, perseverance, dealing with stress, confidence in decisions, etc.

**Notes about Tables S3 and S4: discussion of trends in specific codes.**

A few decisions were less likely to be mentioned in interviews, particularly in certain disciplines. See notes in supplemental table 3 for additional details.

- Decisions 1 and 2 (importance and gaps/opportunities). Mentioned less frequently, particularly where the problem was assigned to the expert (often in engineering or industry) or where the importance was implicit (often in medicine). For example, overall, decision 1 was mentioned in 61% of interviews, but that percentage increased to 94% when medicine and industry were excluded from the tabulation (see table S4).
- Decisions 27 (broader implications), 28 (audience), and 29 (present). Depended on the scope of the project being described and the expert's specific role in it, so these decisions had little relevance in some interviews and fields and were not mentioned. 29 was particularly dependent on field, being mentioned in 68% of interviews overall, but increasing to 94% if medicine and industry were excluded (see table S4).
- Decisions 9 (is the problem solvable). This decision is generally implicit in the fact that the expert picked the problem to describe in the interview (they had decided it was worth solving), but often not mentioned. It was less likely to be coded in interviews in medicine, or other interviews where the problem was assigned to the expert. Often when 9 was mentioned explicitly, it was in the context of deciding not to pursue an approach, or when describing the decision that a specific aspect of a problem or question is not solvable.

- Decision 11 (decompose). The experts relatively seldom discussed decomposition explicitly, likely because it had become such a fundamental and automatic part of their problem-solving process. This was particularly true in medicine, where deciding how to decompose was rarely mentioned, although it is well established (and supported by our informal interviews) as being fundamental to how the medical diagnostic process is structured (e.g. thinking through organ systems). Overall, decision 11 was mentioned in 68% of interviews, but that increased to 76% if medicine were excluded (see table S4), and likely that is still a significant undercount of the true use judging from the informal interviews.

- Decision 17 (represent and organize information), and to a lesser extent 16 (calculations and data analysis). These usually came up explicitly in interviews only if/after the expert was asked how they arrived at conclusions. Without such prompting, the expert would typically describe the information they collected, and then what they interpreted or concluded from that information, without elaborating on how the data was analyzed unless asked. Thus 16 and/or 17 must have happened during this process, but we didn't have enough evidence to code for them. In medicine in particular, 16 and 17 were unlikely to be mentioned, because typically a doctor is provided with test results that are already analyzed by a lab or radiologist, so they do not have to make decisions themselves about how to analyze and represent the data. Overall, 16 was mentioned in 80% of interviews, but that increases to 96% if medicine was excluded (table S4).

- Decisions 18 (how believable is information?) and 20 (any significant anomalies?) were coded less frequently than some other decisions, probably because of the retrospective nature of the interviews. Depending on the problem context, experts may not have encountered significant anomalies or needed to question the validity of information, or they did not come up in the context of the interview, because by that point they had figured out any such behavior, and so in the retrospective process it no longer stood out as puzzling or unexpected.

- Reflection decisions 23 (reflect on assumptions) and 24 (reflect on knowledge) were coded less frequently than others because our coding of "reflection" required the expert to remember and relay their thinking process, in contrast to describing their actions which was the usual focus. In addition, to distinguish 24c from 13 (what information is needed?) and 15 (plan), we required some explicit evidence of reflection, such as statements about re-thinking or deciding to collect additional different information than in original plan.

We also noted some decisions that were particularly likely to co-occur, or in the context of the interview were completely intertwined. Our initial list had several decisions that we found nearly always co-occurred, and those we consolidated. However, some decisions were particularly likely to co-occur but were still mentioned separately a modest fraction of the time, so we kept those separate. We describe the most common of these below and see the additional notes in table 3.

- Decisions 23 (reflect on assumptions), 25 (reflect on strategy), 26 (reflect on solution). It was often difficult to distinguish between reflection decisions. In some cases, the method (approach) *was* the solution to a problem or sub-problem being discussed, so 25 and 26 were identical. In other cases, reflection on solution or approach also required reflecting on assumptions, so 23 often co-occurred with 25 and 26.

- Decisions 3 (goals), 6 (narrow problem), and 11 (decompose). Each is a more specific aspect of refining the problem, and so they could be indistinguishable in an interview, depending on the coherence and detail of the interview.

- Decisions 14 (priorities), 3 (goals), 15 (plan), and 12 (particular difficulty). 14 was often co-coded with other decisions, because deciding on priorities involves the weighting of a variety of factors. Decisions about resources were often coded with 3 (regarding criteria or constraints). Decisions about which approaches to try first or which parts of the problem to approach first were often coded with 15 (plan) or 12 (particular difficulty), because the expert often plans to prioritize (or prioritize ways to avoid) the areas of difficulty identified in 12.

- Decisions 1 (importance), 2 (opportunity fits expertise), and 27 (broader implications). These require very similar cognitive processes, but at different parts of the problem-solving process. Given the structure of our interviews, they were often hard to distinguish. The experts often discussed the importance of the problem as it related to what opportunities there were to make progress that matched with their expertise, so 1 and 2 were frequently coded together. For 27, after discussing their process and solution to a problem, broader implications were often mentioned in the context of discussing their next steps in terms of goals and opportunities, thus 27 would lead to a new problem and a new round of 1 and 2. A subset of interviews had a specific 27 + 2 combination: The expert would describe their development of a new tool or theory, then move on to talking about what problem(s) they could solve using this tool. This also involved 9, in that they were examining what current outstanding problems in the field would now be tractable.

A key feature of the interviews that is not captured by the decisions list is the iterative nature of the decisions being made. Most of the decisions were mentioned multiple times in each interview. Sometimes they were separated into discrete cycles of going through a set of decisions to solve one component of the problem, then repeating to solve a different component of the problem. Often, reflection or an unexpected result or difficulty would trigger iteration back to earlier steps, where the expert would try a different approach at solving the problem or to refine the goals of the problem and solve a modified problem. The experts would also describe problems within problems, for example they would have a bigger-picture problem of trying to answer a scientific question or create a tool, but would also describe in detail the problem-solving process involved in troubleshooting a technical aspect of one of the steps needed in the bigger problem.

**Table S1. (separate file)**

Final coding tabulation of decisions that occurred in each semi-structured interview.

**Table S2. (separate file)**

Notes on complete problem-solving decisions list, and examples of each decision.

**Data S1. (separate file)**

Collection of semi-structured interview transcripts (with redactions for privacy) from subset of experts who agreed to have their interview transcript published. To view, go to LINK XXX TO BE PROVIDED PER LSE GUIDANCE.

**SI References**

1.  Crandall, B., Klein, G. A., Hoffman, R. R. (2006). *Working Minds: A Practitioner's Guide to Cognitive Task Analysis*. MIT Press.